**Title**

On the R2* Relaxometry in Complex Multi-Peak Multi-Echo Chemical Shift-Based Water-Fat Quantification: Applications to the Neuromuscular Diseases

**Authors**

G. Siracusano[1,2*], A. La Corte[2], C. Milazzo[3], G. P. Anastasi[3], G. Finocchio[1,4] and M. Gaeta[3]

[1] Department of Mathematical and Computer Sciences, Physical Sciences and Earth Sciences, University of Messina, V.le F. D'alcontres, 31, 98166 Messina, Italy

[2] Department of Computer Engineering and Telecommunications, University of Catania, Viale Andrea Doria 6, 95125 Catania, Italy

[3] Department of Biomedical sciences, Dental and of Morphological and Functional images, University of Messina, Via Consolare Valeria 1, 98125 Messina, Italy

[4] Istituto Nazionale di Geofisica e Vulcanologia (INGV), Via Vigna Murata 605, 00143 Roma, Italy

* Corresponding Author: giuliosiracusano@gmail.com

**Abstract**

**Purpose:** Investigation of the feasibility of the $R_2^*$ mapping techniques by using latest theoretical models corrected for confounding factors and optimized for signal to noise ratio.

**Theory and Methods:** The improvement of the performance of *state of the art* magnetic resonance imaging (MRI) relaxometry algorithms is challenging because of a non-negligible bias and still unresolved numerical instabilities. Here, $R_2^*$ mapping reconstructions, including complex-fitting with multi-spectral fat-correction by using single-decay and double-decay formulation, are deeply studied in order to investigate and identify optimal configuration parameters and minimize the occurrence of numerical artifacts. The effects of echo number, echo spacing, and fat/water relaxation model type are evaluated through both simulated and in-vivo data. We also explore the stability and feasibility of the fat/water relaxation model by analyzing the impact of high percentage of fat infiltrations and local transverse relaxation differences among biological species.

**Results:** The main limits of the MRI relaxometry are the presence of bias and the occurrence of artifacts which significantly affect its accuracy. Chemical-shift complex $R_2^*$-correct single-decay reconstructions exhibit a large bias in presence of a significant difference in the relaxation rates of fat and water and with fat concentration larger than 30%. We find that for fat-dominated tissues or in patients affected by extensive iron deposition, MRI reconstructions accounting for multi-exponential relaxation time provide accurate $R_2^*$ measurements and are less prone to numerical artifacts.

**Conclusions:** Complex fitting and fat-correction with multi-exponential decay formulation outperforms the conventional single-decay approximation in various diagnostic scenarios. Although it still lacks of numerical stability which requires model enhancement and support from spectroscopy, it offers promising perspectives for the development of relaxometry as a reliable tool to improve tissue characterization and monitoring of neuromuscular disorders.

# 1. Introduction

Potentially, relaxometry [1] offers multiple applications in MRI. Above all, fat/water quantification is considered a promising tool in modern healthcare and the premier non-invasive method for measuring both the amount and the distribution of lipids in biological tissues. Particularly in the last decade, MRI demonstrated its importance as a cost effective solution for diagnosis and monitoring of nonalcoholic fatty liver disease (NAFLD) [2,3,4,5], neuromuscular disorders (NMD) [6] (such as Duchenne Muscular Dystrophy (DMD) [7] and Pompe pathology [8]). To this end, it is well acknowledged that for a correct evaluation of the fat/water percentage it is necessary to take into account the right intrinsic relaxation properties $R_2^*$. As known, $R_2^*$ relaxation includes effects induced by spin-spin relaxation ($R_2$) and by $B_0$ inhomogeneities ($R_2'$), so that $R_2^* = R_2 + R_2'$. Nowadays, transverse relaxation times can be computed in many ways, such as via fast spin-echo (FSE), multiple spin-echo imaging [9,10], driven-equilibrium single-pulse observation time (DESPOT) [11] or spoiled gradient recalled (SPGR) multi-pulse acquisitions. Recent attempts to obtain an $R_2^*$ map considered the use of models which implement voxel-wise [12] or pixel-wise measurements [13]. For a given model, the parameter sensitivity and the reliability in species quantification depend on the repetition time (TR), the slice thickness, the number $N$ of images acquired with different inter-echo spacings (ΔTE), and the macroscopic $B_0$ inhomogeneity [14,15].

Currently, main chemical shift-based approaches can be broadly classified into: magnitude-based [16] and complex-based [4,17]. In the former, phase information is discarded and thus the contribution related to the field map inhomogeneity is not estimated. Therefore, water-fat ambiguity cannot be fully resolved and Fat Fraction (FF) can be uniquely estimated in a 0-50% range [18]. In the latter, the complex-based approach aims to mitigate more confounding effects, preserves the Gaussianity of the noise statistics of MR images [19] and enables measurement of the signal fat-fraction over a theoretical range of 0–100%. To improve the $R_2^*$ measurement in term of robustness, accuracy, precision [16,20,21,] and signal-to-noise ratio (SNR) [22], different numerical formulations have been implemented [23,24]. The most sophisticated ones include a complex-based multi-peak fat signal representation [25] with $R_2^*$ correction that is modeled by using a single (1D) [26] or double decay (2D) approximation for the water and the fat species.

It has been recently demonstrated that 1D complex-fitting model [27] can provide results with high noise stability over a wide range of $R_2^*$ values (0÷600s$^{-1}$) and TE combinations. On the other hand, the accuracy of $R_2^*$ quantification at larger rates [28,29] is currently debated because it is strongly sensitive to the type and timings (ΔTE) of the echo sequence. In addition, relaxometry methods have still notable limitations in obtaining reliable measurements in patients with iron overload in the liver [30,31] and in the myocardium [32,33,34], because of the significant different decay times of water and fat species. Hence, an optimized set of parameters is required to avoid biased reconstructions.

Latest achievements show the possibility to model separate decay rates for water and fat (2D model) [23,35] in order to improve the accuracy of relaxometry mapping and fat quantification [36]. However, they point out how this approach currently suffers of an increased noise sensitivity. For example, it is well known that, the inclusion of fat in tissues can lead to remarkable alterations in the signal amplitude of images acquired at increasing TE [37,38,39,40,41,42]. Although most of such infiltration in organs usually does not exceed 50%, this is not true in degenerative muscle diseases, where higher FF values can be reached up to the complete substitution of the muscular tissue with fat and fibrosis [43,44,45,46].

To this end, $R_2^*$-corrected fat quantification is particularly important in subjects affected by progressive Neuromuscular Disorders (NMD) for the careful assessment of disease severity at the time of diagnosis and for longitudinal monitoring of their response to therapy [47]. The purpose of this study is to analyze the ability of MRI to obtain robust $R_2^*$ quantification by investigating the use of a complex multi-echo chemical-shift based $R_2^*$ estimation method that contemporarily evaluates independent relaxation rates for biological species.

A comparison between 2D and 1D decay models is presented in order to focus on the following challenges: (*i*) reproducibility of results across different FFs, (*ii*) dependence on imaging parameters and protocols, (*iii*) identification of numerical artifacts, (*iv*) requirements to achieve optimal performance. The approach employed concerns to an attempt to extend investigation on previously reported techniques for 1D $R_2^*$ mapping [27] following a systematic approach [48] where the effects of parameter changes are independently investigated. In details, we evaluate the performance of different complex multi-peak fat-corrected relaxometry models in terms of SNR and bias in order to study the limits of each theoretical formulation, such as the weaknesses in $R_2^*$ mapping associated with increasing fat infiltration and for different relaxation rates. In any case, the unknown parameters (e.g. water and fat signal amplitudes, $B_0$ field inhomogeneity, relaxation rates) are evaluated at the same time via the nonlinear least-squares (NLLS) estimation algorithm for simulated and in vivo data. Systematic analyses ranging from the theoretical characterization of different models to simulations and clinically significant examples are presented.

## 2. Methods

*2.1 Theoretical models*

The fat/water quantification from a signal $s_n$ measured on a given voxel at $TE_n$ $(n=1,...,N)$ is achieved by considering a theoretical model. The complex formulation of a single $R_2^*$ decay (1D) including $B_0$ field inhomogeneity is given by:

$$s_n\left(\rho_W,\rho_F,\phi_0,f_B,R_{2C}^*\right) = \left(\rho_W + \rho_F \sum_{p=1}^{P}\alpha_p e^{i2\pi f_{F,p}TE_n}\right)e^{i(\phi_0+2\pi f_B TE_n)}e^{-R_{2C}^* TE_n} + \eta_n \quad (1)$$

where $\rho_W$ and $\rho_F$ are the amplitudes of water and fat signals, respectively, with initial phase $\phi_0$, $f_B$ is the frequency shift due to local magnetic field $B_0$ inhomogeneities, $R_{2C}^* = 1/T_{2C}^*$ is the common decay for both species, while $f_{F,p}$ are the known frequencies for the multiple spectral peaks of the fat signal relative to the water peak and $\alpha_p$ are the relative amplitudes of the fat signal that satisfy the condition $\sum_{p=1}^{P}\alpha_p = 1$. The values of $\alpha_p$ and $f_{F,p}$ can be directly estimated from the data by means of spectrum self-calibration algorithms [2] or well known multipeak spectral configurations [49,50]. It has been recently demonstrated that fat quantification techniques using multipeak fat models [51] provide comparable results. Therefore we will use the method proposed in Ref. [25], using the data as reported in Ref. [49] where the relative amplitudes (%) are $\alpha_p$ = (4.7, 3.9, 0.6, 12, 70, 8.8) whereas the relative frequencies (expressed in ppm) of fat peaks are (0.6, 0.5, 1.95, 2.6, 3.4, 3.8), respectively. The noise $\eta_n$ can be modeled as a complex white Gaussian distribution. Relaxation rate $R_{2C}^*$ can be estimated from Eq. (1) using NLLS that, according to Ref. [52], provides the maximum-likelihood estimation. Above equation assumes a common decay rate for the water and fat signals which has been shown to be effective for fat quantification [12,21,27] in previous phantom [53], and clinical studies [4].

Nonetheless, we stress the fact that a 1D $R_2^*$ model is intuitively not appropriate for an accurate estimation of water and fat compound percentages when they exhibit very different decay rates, such as in the presence of a non negligible iron concentration in muscular tissues [54] or liver [55].

In general, water and fat have different $R_2^*$ decays (and even the multiple fat peaks will have independent decay rates, but this is typically ignored being that it leads to a significant complication of the model [25]). This has led to methods that employ independent $R_2^*$ decay rates [23]. Although multiple $R_2^*$ correction may reduce bias by more accurately modeling the underlying physics, recent studies pointed out the higher numerical instability [25] that requires further investigation. The Eq. (1) can be rewritten for a 2D formulation as follows:

$$s_n\left(\rho_W,\rho_F,\phi_0,f_B,R_{2W}^*,R_{2F}^*\right) = \left(\rho_W e^{-R_{2W}^* TE_n} + \rho_F \sum_{p=1}^{P}\alpha_p e^{i2\pi f_{F,p} TE_n} e^{-R_{2F}^* TE_n}\right) e^{i(\phi_0+2\pi f_B TE_n)} + \eta_n \qquad (2)$$

In this case, independent relaxation rates for water $R_{2W}^*$ and fat $R_{2F}^*$ are considered assuming also the same decay rate $R_{2F}^*$ for all the fat peaks [56]. Thanks to this flexibility, the 2D formulation in Eq. (2) is attractive because it aims to investigate large regional variations of transverse decay rate among biological species. This is particularly appealing in patients developing extensive iron accumulation or severe fat infiltrations in muscles, such as the subjects affected by DMD [57,58] or liver disease [32,53,59]. Besides, the signal may contain transient disturbances due to eddy currents [17] which can be minimized by discarding the phase information of the initial echoes [60] or by constraining the initial phase of both species to be equal at an echo time of zero [61]. To date, the combined effects of 2D relaxometry model and above phase error adjustment method have not been investigated using clinical MRI data for the study of relaxometry and for its application on the NMD assessment. Therefore, we will evaluate complex, multi-peak, 1-decay and 2-decay signal models which also account for phase correction [62]. No field-map smoothing [63,64] or regularization [65,66] techniques are taken into account in this study.

*2.2 Numerical simulations*

Starting from the approach proposed by Hernando et al. [27], we have carried out extensive numerical simulations to characterize the behavior of such theoretical models. The estimated mean, bias and standard deviation (SD) of $R_2^*$ are used to compare the performances of 2D and 1D models. Intuitively, the mean and bias (that quantifies the average difference to be expected between an estimator and the true value of a given parameter) represent an indication of the model mismatch, whereas SD provides information on the precision of the estimator (noise stability of results).

(*i*) First of all, we have analyzed the performance of 2D and 1D methods using a synthetic signal with Gaussian noise having a SNR=50 at TE=0ms and $TE_{init}/\Delta TE$=1.0/1.0ms. Results are proposed in terms of Mean Square Error (MSE) on fitted data as a function of FF.

(*ii*) The robustness and accuracy in $R_2^*$ quantification have been also evaluated by fitting the synthetically generated signals with known values to Eqs. (1) and (2) using both 2D and 1D algorithms. The estimated (EST) relaxation decays under multiple iterations are compared with the 'true' $R_2^*$ data in terms of $bias = \frac{1}{INR}\sum_{j=1}^{INR} true(R_2^*) - EST_j(R_2^*)$. To this end, Monte Carlo simulations have been performed on a total of 1024 independent noise realizations (INR) [67] to provide reliable mean and SD on each reconstruction. Sequences parameters were: *N*=15, SNR=50 at TE=0ms, $TE_{init}/\Delta TE$=1.0/1.0ms.

(*iii*) We have investigated on the error sensitivity of relaxometry mapping as a function of $R_2^*$ value and TE combination using both models with NLLS estimation. Results of including fat-correction in $R_2^*$ quantification are considered in terms of bias and SD. Computations have been performed using different FFs with $R_{2C}^*$, $R_{2W}^*$ and $R_{2F}^*$ ranging within [0-1000s$^{-1}$] under the following configuration: $B_0$=1.5T, $TE_{init}/\Delta TE$=1.0/1.0 ms, SNR=50 at TE=0 ms and INR=1024.

(*iv*) Besides, we have studied the impact of imaging parameters on numerical stability of $R_2^*$ mapping because the quantification of relaxation rates is sensitive to the sequence configuration settings (e.g. initial echo time, echo spacing, echo number) and to the signal model. Theoretical $R_2^*$ noise performance has been investigated by computing mean and standard deviation for 1D ($R_{2C}^*$) and 2D models ($R_{2W}^*$, $R_{2F}^*$), under the following conditions:

- $B_0$=1.5T;
- acquisition using *N*=15 echoes as a function of $TE_{init}$ and $\Delta TE$ within (0-5 ms);
- two different formulations (1D and 2D fat-corrected complex fitting with phase correction method according to Ref. 60);
- 1D model: three relevant rates ($R_{2C}^*$=40, 200 and 1000s$^{-1}$);
- 2D model: independent rates are considered for water-based ($R_{2W}^*$=40, 200 and 1000s$^{-1}$) and fat-based compounds ($R_{2F}^*$=40, 200s$^{-1}$) according to the clinical evidence [68,69,70,71].

(*v*) Further Monte Carlo simulations have been completed to determine the optimal echo number that maximized the dynamic range of relaxometry mapping by testing the bias and noise performance of both decay models using multiple $R_2^*$ combinations with *N*=[9-20] and TE$_{init}$/ΔTE=1.0/1.0ms.

*2.3 Experimental measurements: in vivo validation*

We take into account the numerical robustness of $R_2^*$ quantification in both 1D and 2D fitting models on subjects affected by NMD with increasing levels of fat infiltration in muscles. Robustness has been assessed by reconstructing experimental data from each subject under different echo combinations. A study which includes a cohort of 24 patients is presented. The entire research has been conducted by following the guidelines of the local Ethical Committee whereas a written informed consent has been obtained from each participant. Candidates were selected with definitive diagnoses of DMD, inclusion-body myositis (IBM), Pompe and McArdle disease, and underwent MRI on the pelvic girdle and thigh muscles for the disease clinical assessment between 03/2009 and 02/2015. Part of the patient population included in this study has been also evaluated in prior researches [7,57].

Datasets from the subjects have been acquired on a 1.5T MRI scanner (Gyroscan Intera; Philips, Best, The Netherlands) using a torso phased-array coil. A multiecho SPGR acquisition has been performed with the following parameters: *N*=20/TR=40ms, TE$_{init}$/ΔTE=1.6/1.3ms, FOV=40x40cm, flip angle=5° to minimize T1 residual bias [3], matrix=256x128, NSA=8 and slice thickness=10mm. To test the reliability of reconstructions, datasets have been evaluated using different echo combinations, for *N*=9,12,15,20 echoes. Again, 2D and 1D complex fitting have been adopted with phase correction [60], for a total of eight $R_2^*$ reconstructed mappings per dataset. Measurements have been obtained by an experienced radiologist and compared with [1]H spectroscopy (MRS) data for reference [5].

# 3. Results

*3.1 Simulation study*

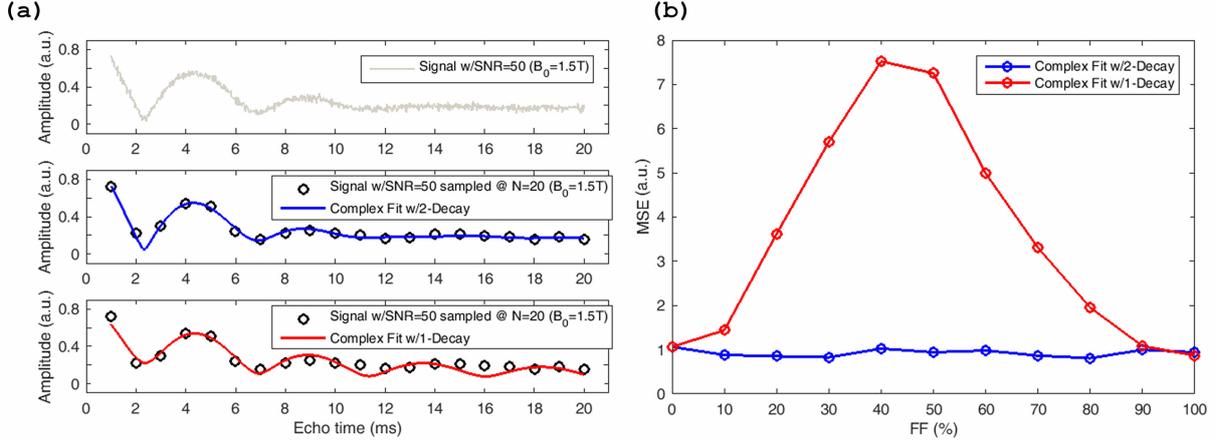

Fig. 1. (a) Synthetically generated 1.5T MRI signal (top gray line) with SNR=50, for FF=50% used as reference and fitted using both 2D (center blue line) and 1D (bottom red line) methods using N=20 echoes (black circles). Reference data are computed by setting water and fat relaxation rates to 200 and 40s$^{-1}$, respectively. (b) Comparison in terms of Mean Square Error (MSE) between chemical shift complex-fitting methods that account for $T_2^*$ decay using double (2D, blue curve) and single decay (1D, red curve) formulation. MSE using 1D model is higher and exhibit a bell-like shape increasing with FF in water-dominated tissues and decreasing with FF in fat-dominated tissues. On the other hand, MSE of 2D model is almost stable among the entire FF range. [Color figure can be viewed in the online issue].

Fig. 1 (a) provides the simulation results of a 1.5T MRI signal (top gray line) with SNR=50, for FF=50%, fitted using both 2D (center blue line) and 1D (bottom red line) methods with a GRE pulse sequences of N=20 echoes (black circles). We impose that $R_{2W}^* = 200 s^{-1}$ and $R_{2F}^* = 40 s^{-1}$ which represent values of interest in skeletal muscles [72]. In this figure, 2D model provides a closer fitting with the samples if compared to 1D formulation. Numerical analysis on simulated data is presented in (b). Here, models performance are compared by computing the MSE between synthetically generated data and their estimated results under varying FF. We impose that water and fat relaxation rates are 200 and 40s$^{-1}$, respectively. MSE using single-decay (1D) model (red curve) is higher and exhibit a bell-like shape which increases with FF in water-dominated tissues and decreases in fat-dominated tissues. On the other hand, MSE of double-decay (2D) model (blue curve) is almost stable among the entire range. Data show how when water and fat will differ in terms of relaxation properties, so these should be estimated separately, leading to the Eq. (2) [23]. In order to discuss noise performance in Fat-Corrected Complex 1D and 2D $R_2^*$ relaxometry, we provide results in Fig. 2.

*3.2 Numerical artifacts*

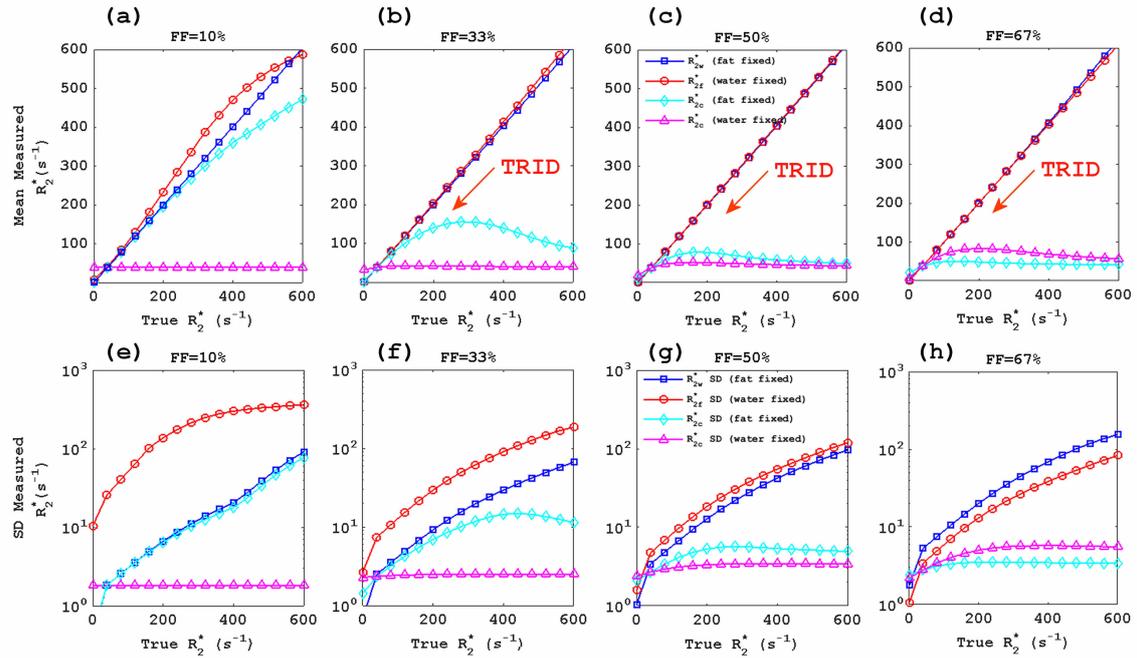

Fig. 2. Comparison of Complex fitting mean (a-d) and standard deviation (e-h) for single and double-decay models using simulated data (INR=1024) from low to severe muscle fat infiltrations. Noise performance is evaluated using 2D model in terms of $R_{2W}^*$ (blue squares) by letting water species vary in reference signal between [0-600s$^{-1}$] whereas fat compounds relaxation rate is set to a common value (40s$^{-1}$), and vice versa ($R_{2F}^*$, red triangles). With the same approach, we analyze 1D numerical stability by investigating on results when water relaxation rate ($R_{2C}^*$, cyan diamonds) varies in reference signal (and fat decay rate is set to 40s$^{-1}$) and vice versa ($R_{2C}^*$, magenta circles). [Color figure can be viewed in the online issue].

Fig. 2 (a-d) plot mean for single and double-decay models using simulated data (INR=1024) from low to severe (67%) muscle fat infiltrations, while (e-h) provide the corresponding standard deviation, respectively. In our simulations, we independently evaluate results using 2D model by letting water species ($R_{2W}^*$, blue squares) vary between [0-600s$^{-1}$] whereas fat compounds relaxation rate is set to a common value (40s$^{-1}$), and vice versa ($R_{2F}^*$, red triangles). With the same approach, we analyze 1D accuracy by studying water ($R_{2C}^*$, cyan diamonds) and fat ($R_{2C}^*$, magenta circles) numerical stability.

It is interesting to note how 1D model exhibits a mismatch in relaxation rate quantification which increases with the FF ratio. We have called this artifact as *Transverse Relaxation rate Inter-species Diversity* (TRID) because its occurrence is originated by the large dissimilarity in terms of decay time between independent chemical species in a given region of interest (ROI). Under the same conditions, 2D model achieves unbiased results up to large rates (600s$^{-1}$). To our best knowledge, previous studies have been performed when the difference between water and fat relaxation as known [16], but they did not consider the scenario where such difference is large, i.e. when the ratio, $\gamma$, between the fastest and the lowest decaying species is, $\gamma \geq 5$. In order to investigate on such findings we have analyzed numerical stability for each variable in Eqs. (1) and (2) by evaluating its bias. As shown in Fig. 3, for most of the clinically relevant fat concentrations, bias curves provide useful insights to enlighten the different behavior between 1D and 2D models. Particularly, such achievements suggest a probable relationship between the offset exhibited using 1D formulation and by the corresponding error in $f_B$ quantification (cyan and magenta curves) if compared with the 2D model (blue and red curves).

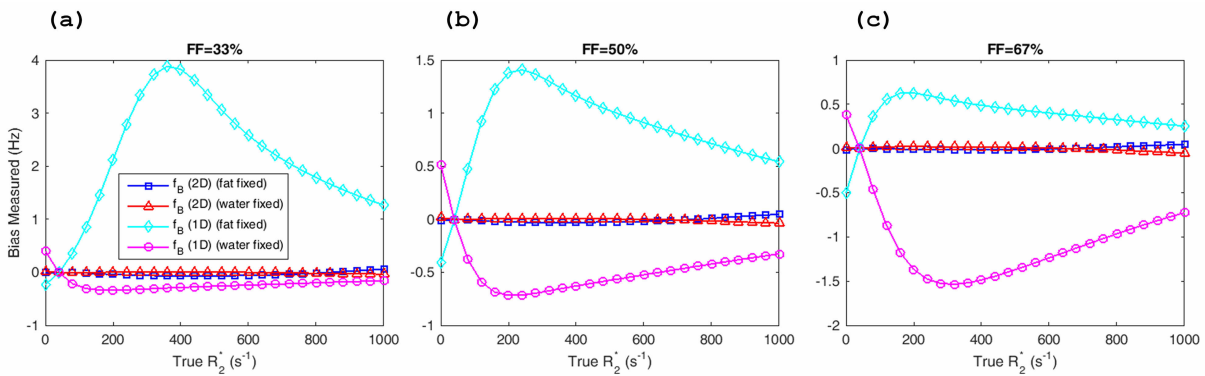

Fig. 3. Comparison of Complex fitting bias using 2D and 1D is proposed (INR=1024, SNR=50) by considering the impact of the estimation problem due to $B_0$ inhomogeneity, $f_B$, on the TRID phenomenon. We evaluate bias in $f_B$ using 2D model by letting water vary in reference signal whereas fat is fixed to 40s$^{-1}$ (blue circles), and vice versa (red triangles) for three different FF ratios (a-c). 1D model performance are shown on each plot in the same way by reproducing $f_B$ bias, as a result of the estimation problem, when water varies in reference signal (cyan diamonds) or fat (magenta circles). [Color figure can be viewed in the online issue].

For moderate (a), mild (b) and severe (c) fat concentrations 2D model shows very stable predictions of $f_B$ for a whole range of relaxation times. Instead, 1D model exhibits a bias which depends on the FF and independently affects the estimation of $R_{2C}^*$. Additional studies have been performed by comparing results of 1D and 2D model when $f_B$ is known. Results (not shown) demonstrate no role of $f_B$ in TRID phenomenon. We believe that such negative aspect of 1D formulation is mainly due to the raw modeling of the dynamical evolution of chemical species by using a single exponential (common relaxation rate), and that it is an intrinsic limit of such model. In summary, when tissues exhibit very different relaxation times, 1D approximation is not recommended due to the observation of a major numerical error which significantly perturbates the quantification of transverse relaxation rate. Again, considering FF ratios in Fig. 2, if we study tissues having non-negligible adipose infiltrations (b-d) using 1D model we might incur in an estimation problem of $R_2^*$ due to the TRID phenomenon which is mainly influenced by the distinct spin-spin interaction properties of real chemical species under investigation. Particularly, it significantly affects results in the presence of considerable regional differences among chemical species which might be attributable to a non-uniform fat distribution in tissues, a common pattern of abnormality related to the disease progression in NMD subjects [73].

*3.3 Numerical stability: comparison between 1D and 2D model*

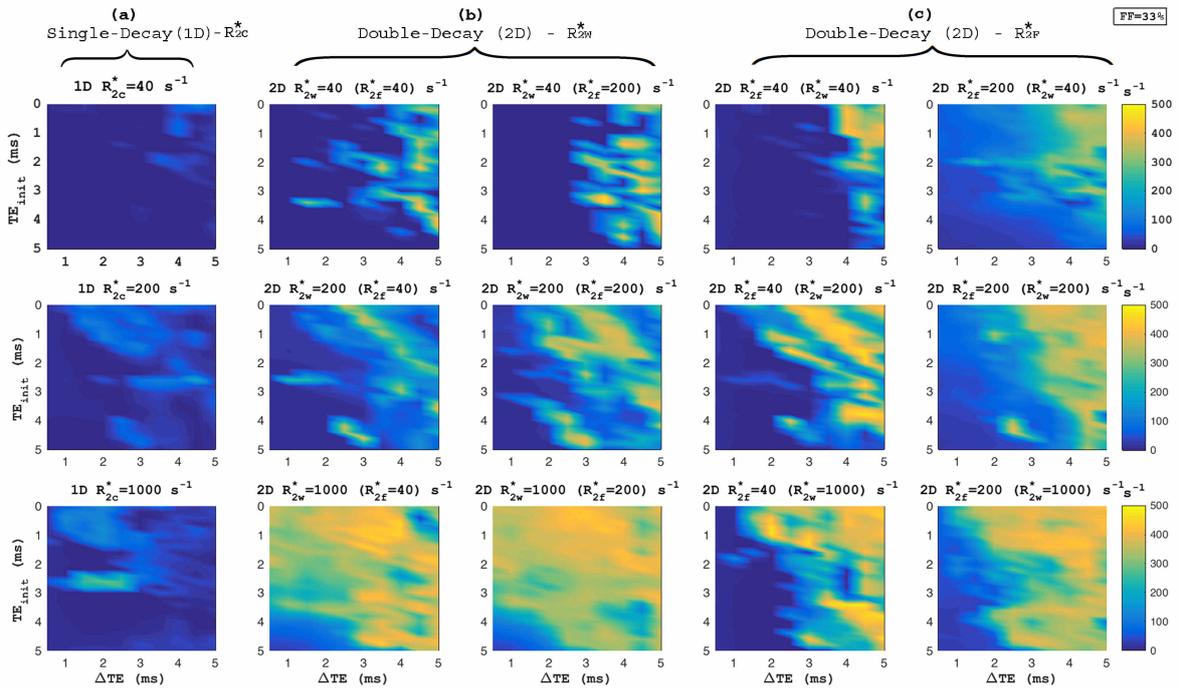

Fig. 4. Theoretical noise performance (standard deviation) of $R_2^*$ estimation at 1.5T for N=15, FF=33% using various TE combinations and different relaxation rates between chemical species. Plots include two distinct reconstruction techniques: (a) Single-decay (1D) and (b, c) Double-decay (2D) complex fitting with multipeak fat spectral representation and phase correction. For 1D formulation (a) $R_{2C}^*$ performance is evaluated under three rates (40, 200, 1000s$^{-1}$). Regarding 2D model, in (b) we estimate noise stability in terms of $R_{2W}^*$ (by letting water relaxation rate vary while fat is set to 40s$^{-1}$) the other parameter, $R_{2F}^*$, is set to the following values [40, 200s$^{-1}$]), and in (c) for $R_{2F}^*$ (when $R_{2W}^*$ is set to [40, 200, 1000s$^{-1}$]). The plots depict the numerical influence of the TE combinations and different relaxation rates between chemical species on the models noise performance. [Color figure can be viewed in the online issue].

Fig. 4 (a-c) are arranged in three sub-matrices and show data for moderate fat (FF=33%) while additional findings are discussed for varying FF in the range (0-100%) but not shown (see Figs. 4.2-4.6 in supplemental materials). Each matrix plots the SNR performance (standard deviation) for a given signal model as a function of $TE_{init}$ and $\Delta TE$ under different rate combinations. We calculate the SD of $R_{2C}^*$ (a) for single-decay model, and of $R_{2W}^*$ (b) and $R_{2F}^*$ (c) for double-decay model, respectively. 1D model results are evaluated by considering three rates (40, 200, 1000s$^{-1}$) for water whereas fat is set to (40s$^{-1}$). For the 2D model, in (b) we estimate noise performance for varying $R_{2W}^*$ (while the other parameter, $R_{2F}^*$, is set to the following values [40, 200s$^{-1}$]), and in (c) for $R_{2F}^*$ (when $R_{2W}^*$ is set to [40, 200, 1000s$^{-1}$]). On one side, shortened echoes ($\Delta TE<1.5$ms) in 1D model provide the best noise performance from low (40s$^{-1}$) to very high relaxation rates (1000s$^{-1}$). For increasing FF, numerical stability of $R_{2C}^*$ increases. On the other side, from low to moderate relaxation rates (0-200s$^{-1}$) 2D model exhibits different performance between $R_{2W}^*$ and $R_{2F}^*$ estimation. Our results show that $R_{2W}^*$ and $R_{2F}^*$ numerical stability is higher for $\Delta TE<1.5$ms [74]. In addition, when water relaxes rapidly (1000s$^{-1}$) the numerical instability causes a sensible decrease of SNR. This can be mainly associated with the difficult rate estimation of water compound due to the immediate exponential loss of signal strength after few milliseconds. For increasing FF, noise performance decreases in 2D model either for $R_{2W}^*$ and, less markedly, for $R_{2F}^*$. In water-dominated tissues and in the presence of very high decay rates, simulation data highlight how short echoes are fundamental for reliable rate predictions for 1D and 2D formulations. In this scenario, $R_{2C}^*$ generally exhibits a better noise stability (lower SD). Interestingly, when fat is predominant and for a rapid water decay ($R_{2W}^*$ = 1000s$^{-1}$) 2D model shows an apparent reduction of SD near $\Delta TE\approx4.5$ms which can be explained as the consequence of the combined effect of: (*i*) loss of signal information caused by the rapid decay of water, (*ii*) large inter-echo time ($\Delta TE$) which is similar to in-phase (IP) acquisitions that have been commonly used to avoid the effect of fat in $R_2^*$ mapping [75]. It is interesting to note how TRID phenomenon is not evident in the above figure, because it is a prerogative of 1D model.

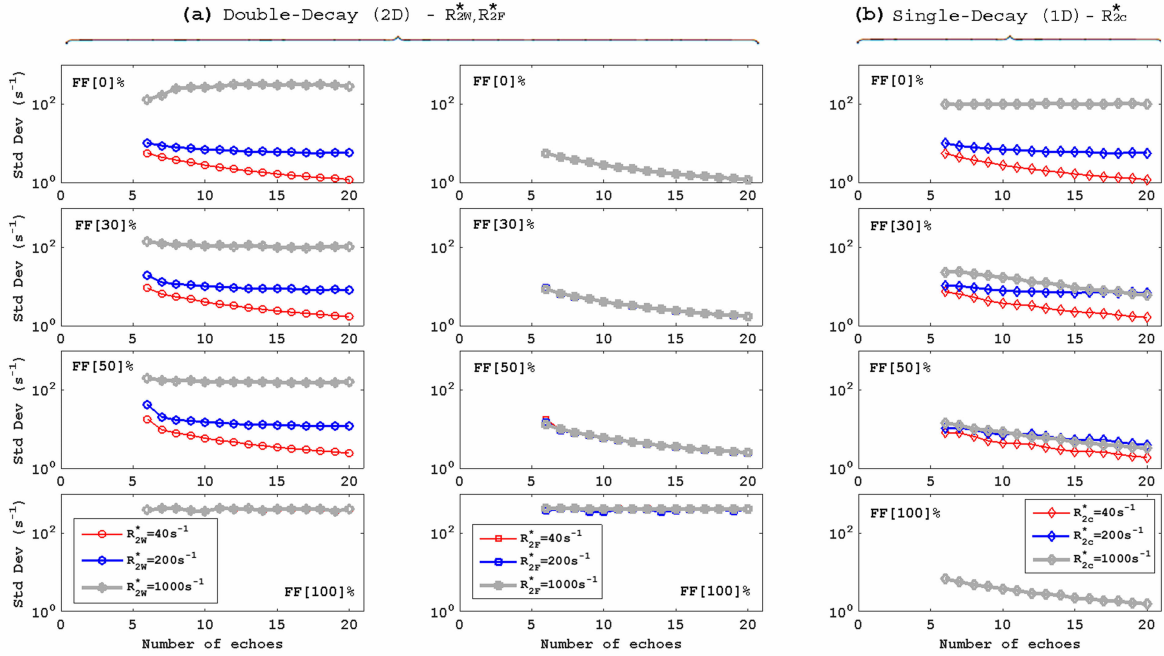

Fig. 5. 2D and 1D $R_2^*$ mapping as a function of $N$ and FF (SNR=50, INR=1024). Plots show the theoretical noise performance (standard deviation) under different TEs, for fixed initial TE (TE$_{init}$=1 ms) and echo spacing (ΔTE=1 ms). The optimum number of echoes for fat-corrected $R_2^*$ mapping heavily depends on the model used and on the relaxation rate of the biological compounds. The minimum echo train length recommended to support both formulation is 9, although N≥15 is considered to be optimal for 2D model. [Color figure can be viewed in the online issue].

Here, we investigate the effects of the number of echoes $N$ acquired during imaging, for different relaxation rates. Fig. 5 (a),(b) show the theoretical noise performance (SD) of complex fat-corrected $R_2^*$ mapping for TE$_{init}$/ΔTE=1/1ms as a function of $N$ (6-20) and FF (0-100%). Noise performance is analyzed by considering 1D and 2D model (INR=1024). In this figure, the rows depict results for a given FF while columns represent SD for (a) water $R_{2W}^*$, fat, $R_{2F}^*$ (in 2D model using Eq. (2)), and (b) common $R_{2C}^*$ (in 1D model using Eq. (1)), respectively. We highlight that for $R_{2W}^*$ ($R_{2F}^*$) results are considered by setting complementary fat (water) relaxation rate to a reference value (40s$^{-1}$), whereas for $R_{2C}^*$ we set both species to the same rate. For the sake of completeness, numerical stability is evaluated using three known decay values: low 40s$^{-1}$ (red line), moderate 200s$^{-1}$ (blue line), high 1000s$^{-1}$ (gray line). Again, we point out that clinically relevant relaxation rates for fat are within (0-200s$^{-1}$) [45]. In water-dominated tissues (FF<50%) and from low to moderate decays (<=200s$^{-1}$), the SD for water $R_{2W}^*$ is close to $R_{2C}^*$ and exhibits just a slight reduction when $N$ increases. (b) Although in 1D formulation a value of $N = 6\div8$ is believed appropriate in most applications [27], we observe that the number of 15 is the minimum recommendation for 2D model in order to achieve a better noise performance. As expected, for very rapid decay rates (1000s$^{-1}$), the SD of $R_{2W}^*$ and $R_{2F}^*$ in 2D model is higher than $R_{2C}^*$ in 1D model. This is a predictable result because the relaxation rate is evaluated independently on each species and, when water signal decays very rapidly, it causes an estimation problem in the $R_{2W}^*$ quantification of Eq. (2) due to the lack of available enough information from the echo sequence. In general, such variability is compensated by a reduced bias and an improved accuracy. For a balanced concentration of fat and water compounds (FF=50%), a value which is particularly relevant in the study of NMD [7,57], results indicate that for $N \geq 15$ and from low to moderate decay rates, the standard deviation in the 2D and 1D case have similar values. Interestingly, for rapid decay rates (1000s$^{-1}$) and in fat-dominated voxels, simulations data show that the numerical robustness of 1D model is higher than in 2D formulation.

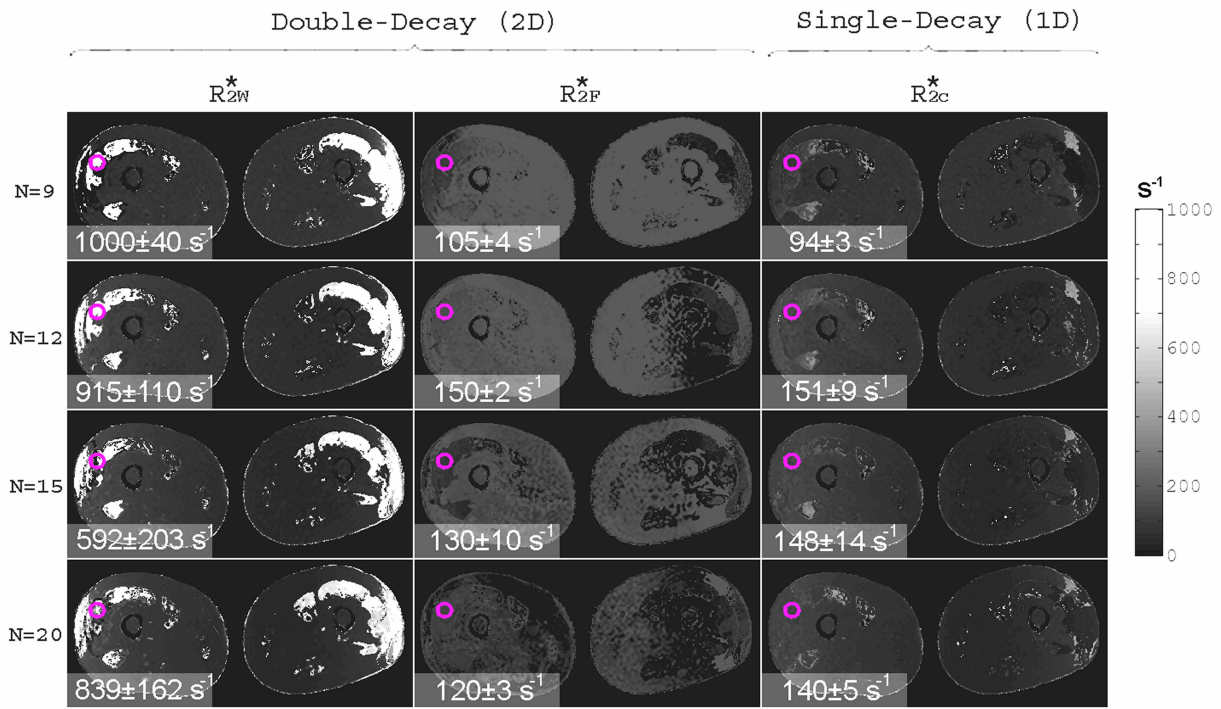

Fig. 6. Fat-correction using multipeak fat modeling is necessary for robust $R_2^*$ mapping in the presence of fat. Images show $R_2^*$ maps of thigh muscles on a subject affected by DMD with low fatty infiltration (FF = 20%), using two different techniques (2D and 1D fat-corrected with multipeak fat) and five different TE combinations (N=9, 12, 15 and 20 echoes). Note the decrease in $R_2^*$ mapping noise for increasing number of echoes. 2D $R_2^*$ measurements show increased accuracy but stronger variability with echo combination, whereas 1D $R_2^*$ estimates demonstrates excellent robustness to echo combination. [Color figure can be viewed in the online issue].

## 3.1 Experimental results: in vivo validation

Fig. 6 provides $R_2^*$ reconstructions in a Pompe patient with iron overload and moderate muscle fat infiltration (MRS PDFF=32%). Results are proposed in terms of bias and SD. Two different fitting methods were compared: complex multipeak fat model with 2D (left) and 1D $R_2^*$-correction (right), respectively. Each reconstruction has been computed with $TE_{init}/\Delta TE$=1.6/1.3ms and evaluated using $N$=[9,12,15,20] echoes. A ROI has been circled to show the results obtained by increasing $N$. Our achievements suggest how 2D Complex Fit is more accurate in the $R_2^*$ mapping of fatty infiltrations in muscles as well as in the presence of iron accumulation. Such results are qualitatively and quantitatively in agreement with experimental findings as shown in Fig. 5. For this analysis, $N$=9 has been evaluated as the minimum number of echoes suitable to provide reproducible results for both 1D and 2D models. However, it is suboptimal for 2D modeling due to the increased complexity of the system which leads to numerical variability in fat-corrected reconstructions. Using 2D model and for moderate or severe fat accumulation in tissues $N\geq15$ is recommended for a better SNR.

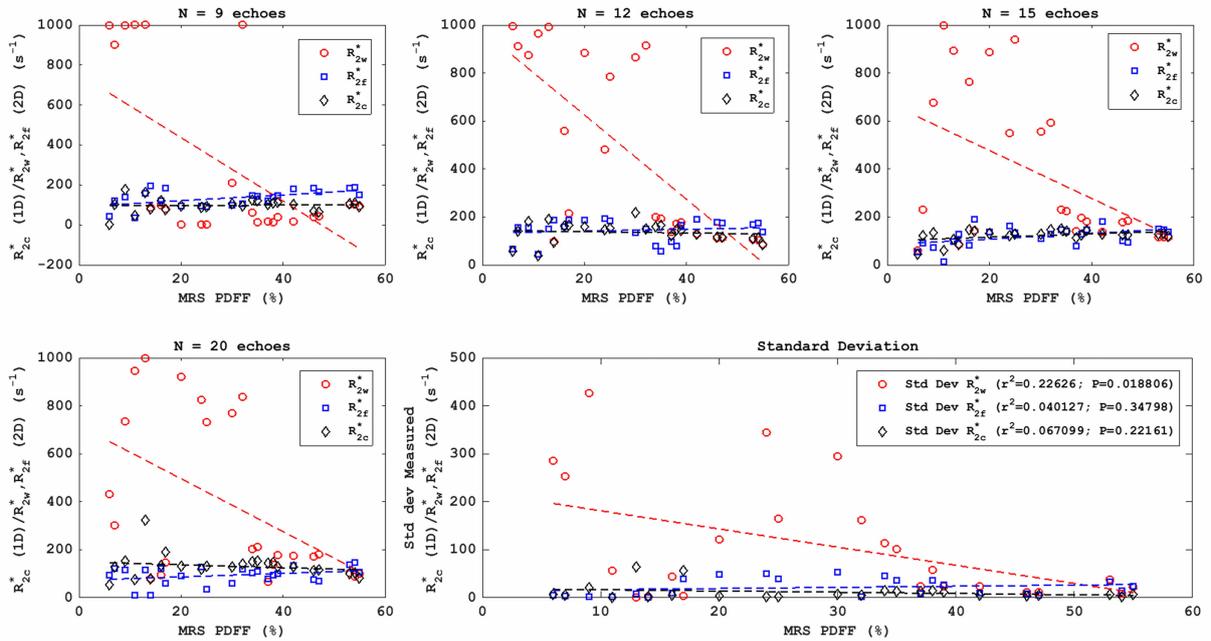

Fig. 7. Fat-correction using multipeak fat modeling is necessary for robust $R_2^*$ mapping in the presence of fat. A cohort of 24 patients affected by NMD has been screened and monitored during this study. (Top) $R_2^*$ measurements as a function of PDFF obtained using 2D and 1D techniques, using four different TE combinations (9, 12, 15 and 20 echoes) . (Bottom) Plot shows $R_2^*$ variability (standard deviation) for each reconstruction, as a function of PDFF. 1D-model (single-decay) reconstructions result in $R_2^*$ variability decreasing with PDFF. 2D-model (double-decay)

$R_2^*$ mapping theoretically should result in numerical variability decreasing with PDFF. In practice, 1D model results in low variability (good robustness) over the entire range of PDFF. [Color figure can be viewed in the online issue].

Fig. 7 exhibits noise stability results from co-localized relaxation measurements obtained from all the subjects for varying numbers of echoes $N$=9, 12, 15, and 20. To avoid deviation of the fitting algorithm far from the biologically likely solutions, $R_{2C}^*$ and $R_{2W}^*$ have been constrained to be greater than 0s$^{-1}$ and not higher than 1000s$^{-1}$, while $R_{2F}^*$ has been limited within (0-200s$^{-1}$). At low fat-fractions and particularly for high relaxation rate in tissues (e.g. heavy iron accumulation), methods perform differently. 2D model is enough sensitive to capture the intrinsic differences in decay between water and fat species, with a stability that increases with $N$, while 1D model reports a low varying relaxation rate which is a far approximation. Robustness is improved in 2D considering $N\geq15$ while single-decay fat-corrected modeling remains substantially stable for $N\geq6$. Multipeak complex fat modeling with single-decay (1D) results in the best robustness (reduced SD), particularly for low to moderate rate values. Such results are in strict agreement with previous findings [25]. Linear regression analysis revealed no statistically significant relationship between the SD of relaxation rates $R_{2W}^*$ ($r^2$=0.22, $P$=0.01), $R_{2F}^*$ ($r^2$=0.04, $P$=0.34), $R_{2C}^*$ ($r^2$=0.06, $P$=0.22) and the fat-fraction, respectively.

## 4. Discussion

The potential of $R_2^*$ mapping methods is extremely attractive for the diagnosis of several pathologies and is a valuable tool for the continuous monitoring of progressive neuromuscular diseases. It aims to address the invasiveness to which traditional biopsy is prone [7] and can contribute to provide robust PDFF quantification. However, the reliability of relaxometry measurements need to be completely understood to gain wide acceptance. In order to obtain reproducible and low bias estimations, we analyzed state-of-the-art fat-corrected $R_2^*$ mapping methods using common (1D) or independent decays (2D). Recent advances pointed out that the most known factors affecting the accuracy, sensitivity, and reproducibility are the imaging parameters, the relaxation properties of tissues and the signal model. For what concerns acquisition parameters, shortened image acquisition times (TE$_{init}$/ΔTE=1.6/1.3ms or less) generally offer significant correlation between estimated relaxation maps and MRS measurements [27], while they are compliant with most of the current MRI appliance. Here, we studied the necessary conditions to improve $R_2^*$ quantification to analyze tissues having very different $R_2^*$ decay rates of water and fat compounds [55,76], and abnormal fat infiltrations. Latest enhancements of 1D and 2D models have been taken into account, such as the adoption of a multi-peak fat signal representation [77] with phase correction [60], self-calibration [2]. Analysis using the 2D model is more accurate (reduced bias) than 1D, and more appropriate to deal with different transverse relaxation properties [78] for water and fat, especially for tissues having non-negligible FF. This comes at the expenses of a lower noise stability which requires further model enhancements. Simulation data using 1D approach carefully reproduce recent achievements and indicate that this formulation is able to maintain the better SNR performance over a broad range of $R_2^*$ values and TE combinations. Nonetheless, it is not widely recommended due to the occurrence of TRID phenomenon, and generally it provides sub-optimal results in scenarios with moderate to severe fat infiltrations. Particularly, it may suffer of a higher bias than 2D model, especially from moderate to critical fat concentrations (FF>30%), which are relatively common physiological conditions in long-term patients affected by NMD. In addition, under some conditions, our results highlight also that measurements of these biomarkers can be significantly altered by biological effects. Specifically, an hallmark feature of 1D model is that the bias increases with the difference in relaxation rates between water and fat species. Such numerical artifact, TRID, represents a significant confounding factor which has been isolated in 1D model whereas it does not affect 2D formulation.

We find that the most favorable combination of acquisition parameters depends also on fat concentration and intrinsic relaxation properties of chemical species. For high decay rates, short TEs are fundamental to minimize bias and standard deviation. In water(fat)-dominated tissues, $R_2^*$ quantification can be improved on 1D method by using a rapid TE$_{init}$ [32] and small (long) ΔTE with *N*≥9 whereas the best results with 2D model can be achieved for *N*≥15. In single-decay formulation, the occurrence of the TRID effect gives new insights in the understanding of relaxometry and aims to highlight some of the limits related to such approximation.

On the other hand, 2D model provides more accurate and TRID-free estimations, but improvements are necessary to reduce its high noise sensitivity. In vivo acquisitions have been performed using a low flip angle (5°) to minimize T$_1$ effects in fat quantification. We do not expect T$_1$ effects to directly introduce bias on $R_2^*$ mapping. Simulations and in vivo results are consistent, and they demonstrate that $R_2^*$ mapping can provide predictable performance for the measurement of the relaxation rates and, consequently, for the FF ratio.

## 5. Conclusions

This article discusses a comparative study of performances of the Multi-Peak Chemical-Shift models with $R_2^*$-correction under single and double-decay approximation. We have performed an in-depth analysis to understand the current limits of *state of the art* techniques and how they could be optimized to increase numerical stability. Such aim is accomplished by isolating the effects of: bias, noise, FF, number of echoes, *N*, echo spacing, relaxation rate variability among biological species.

Our results have pointed out some remarkable characteristics and drawbacks. By considering such experimental findings and taking into account the performance of most 1.5T MRI scanners, we argue that the combination of $TE_{init}/\Delta TE=1/1$ms and $N=15$ is the minimum recommended configuration suitable for both 2D and 1D models and able to conciliate between expected performance and exam duration.

Among the above discussed Multi-Peak Chemical-Shift models the double decay formulation is considered to be a preferable choice for the following reasons: (*i*) it is both theoretically and experimentally more appropriate to deal in scenarios when chemical species exhibit significantly different decay rates, for example, it is particularly appealing in patients developing iron overload or being affected by NMD with severe fatty infiltrations in tissues. (*ii*) Despite 2D formulation demonstrated higher noise sensitivity than 1D model, it provides a lower bias and opens perspectives for future improvements in order to overcome its currently limited noise sensitivity. (*iii*) In addition, it intrinsically corrects for TRID artifact by evaluating both fat and water percentage separately and it might in the future deliver a more reliable and long-term solution for the systematic management of NMD patients. In conclusion, the present work has provided elements to enhance MRI relaxometry by choosing optimal imaging parameters for a given expected range of $R_2^*$ values, while progress is still needed to promote a universal standardization of the relaxometry techniques as a part of emerging MRI diagnostic tools. In fat-dominated tissues and for the monitoring of NMD, 2D fitting is necessary to achieve more consistent $R_2^*$ mapping. Such results have been performed by means of a parallel processing framework which was developed for accelerating algorithms computation and recently demonstrated successful results in other fields [79,80]. Overall, we believe that the conclusions drawn by this study might be seamlessly extended to 3T and to different medical applications where a large transverse relaxation diversity among chemical species is still appreciable, although further work is needed to confirm this speculation. Finally, since fat-quantification is now considered as a necessary measure to perform in MRI studies of all the patients with degenerative muscular diseases [81], we believe that the approach described here contributes to a better understanding on this non invasive technique and thus draw the guidelines to stimulate a desirable progress in future implementations. In the prospect of healthcare applications, such methods could be extensively applied on screening programs to track the evolution of NMD.

**ACKNOWLEDGMENTS**

The simulations have been performed with the availability of SCL (Scientific Computing Laboratory) of the University of Messina, Italy. G. S. acknowledges project entitled "Tecniche innovative di processamento di segnali per lo sviluppo di sistemi e servizi ICT". The authors acknowledge Prof. Placido Bramanti for his support in this project.

**Supplemental Materials - Caption Descriptions**

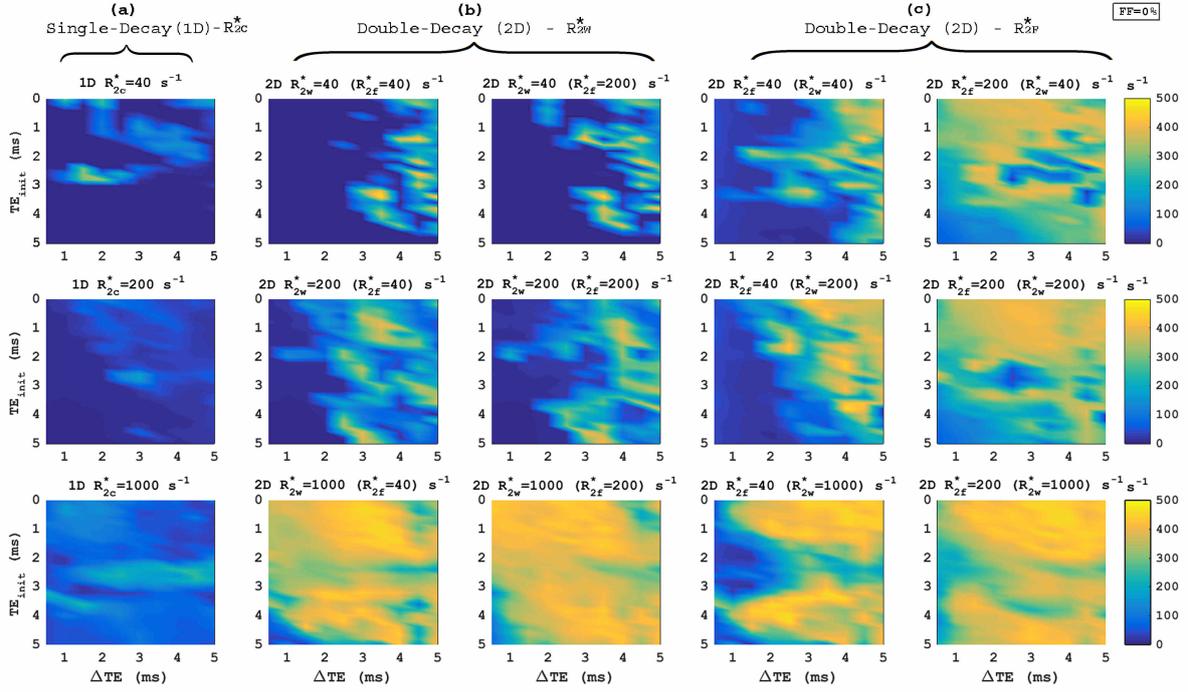

Fig. 4.2. Theoretical noise performance (standard deviation) of $R_2^*$ estimation at 1.5T for *N*=15, FF=0% using various TE combinations and different relaxation rates between chemical species. Plots include two distinct reconstruction techniques: (a) Single-decay (1D) and (b, c) Double-decay (2D) complex fitting with multipeak fat spectral representation and phase correction. For 1D formulation (a) $R_{2C}^*$ performance is evaluated under three rates (40, 200, 1000s$^{-1}$). Regarding 2D model, in (b) we estimate noise stability in terms of $R_{2W}^*$ (by letting water relaxation rate vary while fat is set to 40s$^{-1}$) the other parameter, $R_{2F}^*$, is set to the following values [40, 200s$^{-1}$]), and in (c) for $R_{2F}^*$ (when $R_{2W}^*$ is set to [40, 200, 1000s$^{-1}$]). The plots depict the numerical influence of the TE combinations and different relaxation rates between chemical species on the models noise performance in the absence of fat. [Color figure can be viewed in the online issue].

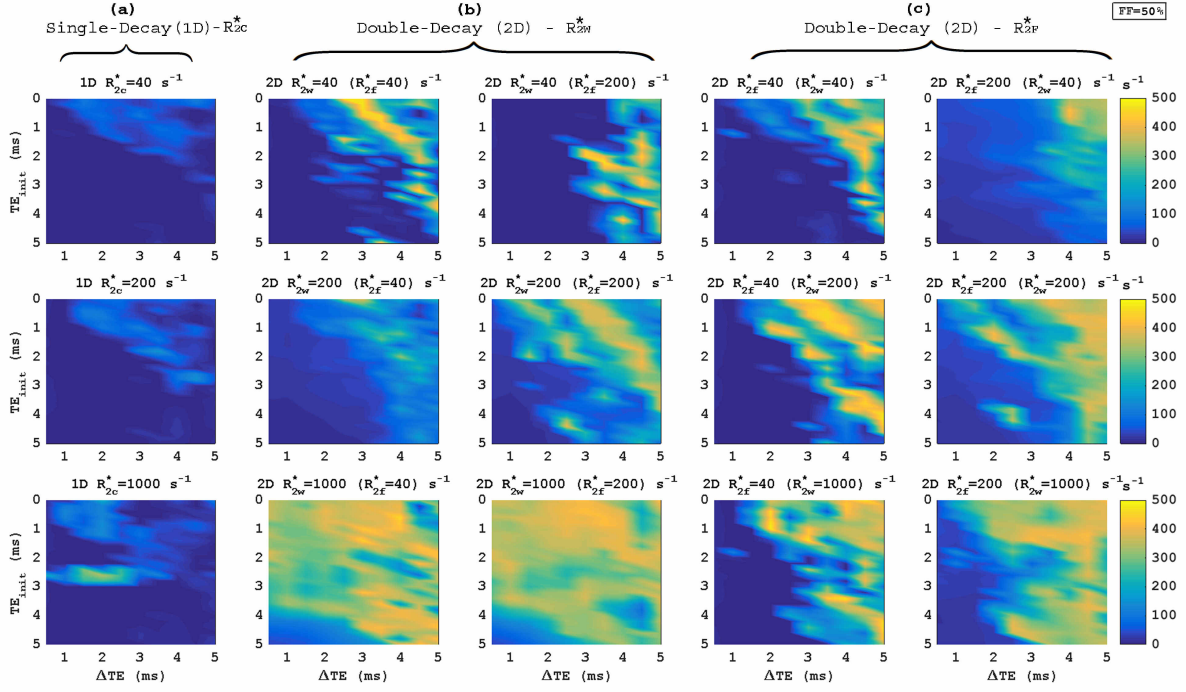

Fig. 4.3 Theoretical noise performance (standard deviation) of $R_2^*$ estimation at 1.5T for $N$=15, FF=50% using various TE combinations and different relaxation rates between chemical species. Plots include two distinct reconstruction techniques: (a) Single-decay (1D) and (b, c) Double-decay (2D) complex fitting with multipeak fat spectral representation and phase correction. For 1D formulation (a) $R_{2C}^*$ performance is evaluated under three rates (40, 200, 1000s$^{-1}$). Regarding 2D model, in (b) we estimate noise stability in terms of $R_{2W}^*$ (by letting water relaxation rate vary while fat is set to 40s$^{-1}$) the other parameter, $R_{2F}^*$, is set to the following values [40, 200s$^{-1}$]), and in (c) for $R_{2F}^*$ (when $R_{2W}^*$ is set to [40, 200, 1000s$^{-1}$]). The plots depict the numerical influence of the TE combinations and different relaxation rates between chemical species on the models noise performance for a balanced ratio water and fat. [Color figure can be viewed in the online issue].

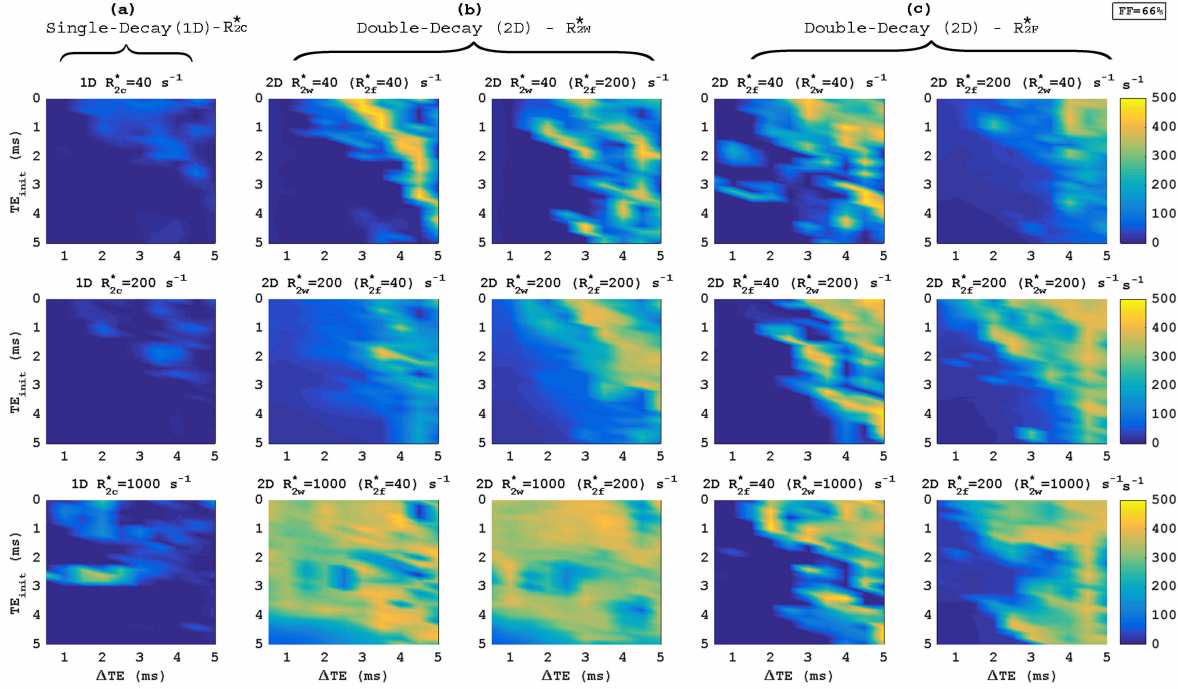

Fig. 4.4 Theoretical noise performance (standard deviation) of $R_2^*$ estimation at 1.5T for $N$=15, FF=66% using various TE combinations and different relaxation rates between chemical species. Plots include two distinct reconstruction techniques: (a) Single-decay (1D) and (b, c) Double-decay (2D) complex fitting with multipeak fat spectral representation and phase correction. For 1D formulation (a) $R_{2C}^*$ performance is evaluated under three rates (40, 200, 1000s$^{-1}$). Regarding 2D model, in (b) we estimate noise stability in terms of $R_{2W}^*$ (by letting water relaxation rate vary while fat is set to 40s$^{-1}$) the other parameter, $R_{2F}^*$, is set to the following values [40, 200s$^{-1}$]), and in (c) for $R_{2F}^*$ (when $R_{2W}^*$ is set to [40, 200, 1000s$^{-1}$]). The plots depict the numerical influence of the TE combinations and different relaxation rates between chemical species on the models noise performance for fat-dominated tissues. [Color figure can be viewed in the online issue].

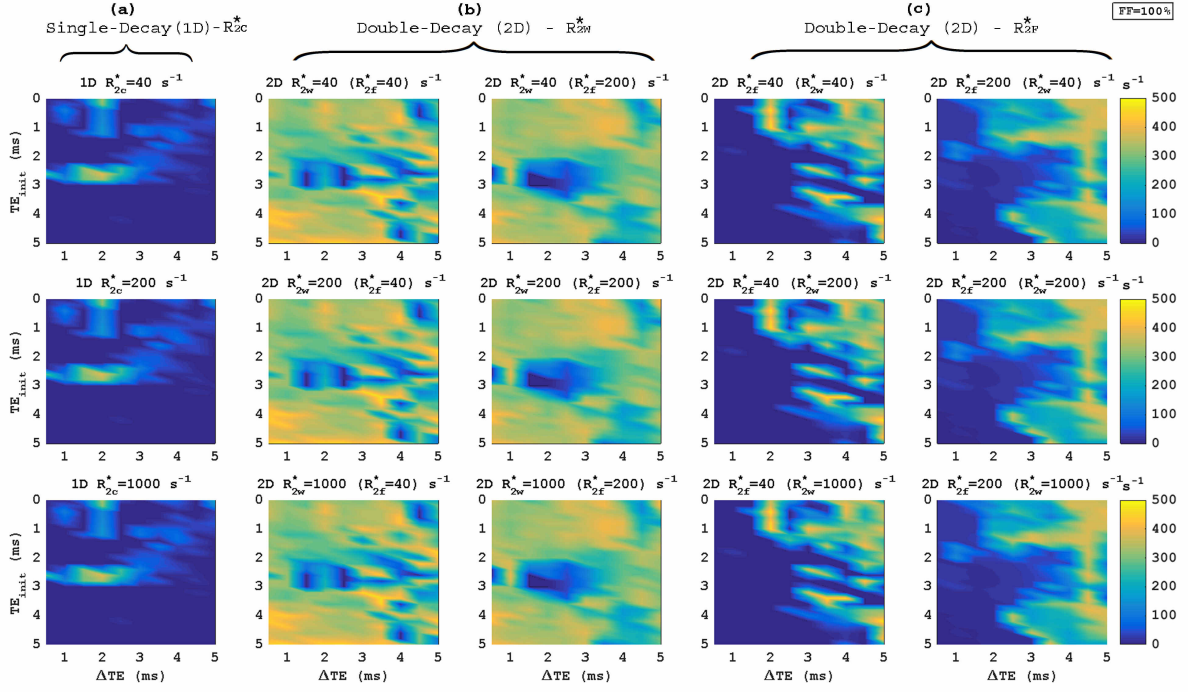

Fig. 4.5 Theoretical noise performance (standard deviation) of $R_2^*$ estimation at 1.5T for $N$=15, FF=100% using various TE combinations and different relaxation rates between chemical species. Plots include two distinct reconstruction techniques: (a) Single-decay (1D) and (b, c) Double-decay (2D) complex fitting with multipeak fat spectral representation and phase correction. For 1D formulation (a) $R_{2C}^*$ performance is evaluated under three rates (40, 200, 1000s$^{-1}$). Regarding 2D model, in (b) we estimate noise stability in terms of $R_{2W}^*$ (by letting water relaxation rate vary while fat is set to 40s$^{-1}$) the other parameter, $R_{2F}^*$, is set to the following values [40, 200s$^{-1}$]), and in (c) for $R_{2F}^*$ (when $R_{2W}^*$ is set to [40, 200, 1000s$^{-1}$]). The plots depict the numerical influence of the TE combinations and different relaxation rates between chemical species on the models noise performance in the absence of water. [Color figure can be viewed in the online issue].